\documentstyle[11pt,aaspp4]{article}
\slugcomment{Submitted to the Astrophysical Journal}

\def\grb{GRB\thinspace{970111}}

\begin{document}

\title{Radio Monitoring of the January 11, 1997 Gamma-Ray Burst}

\author{D. A. Frail\altaffilmark{1}, 
S. R. Kulkarni\altaffilmark{2}, 
E. Costa\altaffilmark{3}, 
F. Frontera\altaffilmark{4,5},
J. Heise\altaffilmark{6}, 
M. Feroci\altaffilmark{3},
L. Piro\altaffilmark{3}, 
D. Dal Fiume\altaffilmark{5},
L. Nicastro\altaffilmark{5}, 
E. Palazzi\altaffilmark{5}, 
R. Jager\altaffilmark{6}}
 
\altaffiltext{1}{National Radio Astronomy Observatory, Socorro,
           NM, 87801, USA} 

\altaffiltext{2}{Division of Physics,
Mathematics and Astronomy 105-24, Caltech, Pasadena CA 91125, USA}

\altaffiltext{3}{Istituto di Astrofisica Spaziale, CNR, Via E. Fermi,
21, 00044 Frascati, Italy} 

\altaffiltext{4}{Dipartimento di Fisica,
Universita' di Ferrara, Via Paradiso 12, I-44100 Ferrara, Italy}

\altaffiltext{5}{Istituto Tecnologie e Studio Radiazioni
Extraterrestri, CNR, Via P. Gobetti 101, I-40129 Bologna, Italy}

\altaffiltext{6}{Space Research Organization, Netherlands,
Sorbonnelaan 2, 3584 CA Utrecht, The Netherlands}

\begin{abstract}

We report on a comprehensive radio monitoring program of the bright
gamma-ray burster \grb. These VLA observations were made at a
frequency of 1.4 GHz ($\lambda=20$ cm) and span a range of post-burst
timescales between 28 hours and one month. Despite extensive sampling
at sub-milliJansky sensitivities, no radio source was detected above
0.5 mJy in the current best error box ($\sim$14 arcmin$^{2}$) for
\grb. A highly unusual radio source, VLA J1528.7+1945, was seen to
drop in flux density by a factor of two in our monitoring period but
it lies outside the error box and thus it is unlikely to be related to
\grb. Cosmological fireball models of gamma-ray bursts make
predictions of late-time emission occurring at longer wavelengths. The
absence of a flaring or fading radio counterpart to \grb\ provides
strong constraints on these models.

\end{abstract}


\vfill\eject
\section{Introduction}

The identity of gamma-ray bursters (GRBs) has remained a mystery since
their discovery more than three decades ago (Klebesadel, Strong \&
Olson 1973). These brief but intense flashes of gamma-rays are
isotropically distributed on the sky, but the distribution of burst
brightnesses appears to be inhomogeneous, in particular showing a
deficit of fainter events (Meegan et al. 1992). The existing data
cannot unambiguously distinguish whether GRBs originate in a halo
around our Galaxy (Lamb 1995) or at cosmological distances
(Paczy\'nski 1995).

The detection and identification of a GRB counterpart at other
wavelengths has the potential to make significant progress in our
understanding of the origin of GRBs. Cosmological models predict that
a radio outburst will follow the gamma-ray outburst (Paczy\'nski \&
Rhoads 1993, M\'esz\'aros \& Rees 1997), albeit delayed and with decay
timescales of order a few days to weeks. There have been previous
searches for radio counterparts of GRBs (e.g. Schaefer et al. 1989,
Palmer et al. 1995, Dessenne et al. 1996).  but none so far have
presented a serious challenge to predictions of the cosmological
models. The principal difficulty has been in obtaining sensitive radio
observations over the full range of the relevant timescales. The
observations of GRB\thinspace{940301} were the first to continuously
sample a range of post-burst timescales between 3 and 15 days, as well
as 26, 47 and 99 days (Frail et al. 1994), but only to a limit of 3.5
mJy at $\lambda=20$ cm.

With the launch of Satellite per Astronomia a raggi X or BeppoSAX on
April 30 1996 it became possible to use the Very Large Array (VLA) to
do a full monitoring of events at high sensitivity over a range of
post-burst timescales.  The gamma-ray burst experiment on BeppoSAX
consists of two parts: the Gamma-Ray Burst Monitor (Frontera 1997)
(60-600 keV) and the Wide Field Cameras (2-30 keV) (Jager et
al. 1996), and is capable of detecting approximately 5$\pm$2 bursts
per year with f$_\gamma>10^{-6}$ erg cm$^{-2}$ (40-600 keV) and
localizing them with a position error of several arcminutes (Piro
1997). Here we present the results of an extensive VLA monitoring
campaign of \grb, in which observations were made beginning 28 hours
after the burst and continued every several days for up to a month.
Prior to these observations the earliest response to a well-localized
GRB at the VLA was 9 days (GRB\thinspace{930706}) by Palmer et
al. (1995). Furthermore, the sensitivity of our VLA campaign is an
order of magnitude improvement over the earlier monitoring effort of
GRB\thinspace{940301} by Frail et al. (1994) on these same timescales.

\section{Observations}

\grb\ was detected (Costa et al. 1997a) by the Italian satellite
BeppoSAX on January 11.41, 1997 UT. It was seen by both the Gamma-Ray
Burst Monitor, and was in the $40^\circ\times 40^\circ$ field of view
Wide Field Camera. The gamma-ray burst had a peak flux of $4-5\times
10^{-6}$ erg cm$^{-2}$ s$^{-1}$ and lasted for a period of about 50
s. Detection of the burst in the WFC enabled an initial localization
of the burst to a circle of radius 10 arcmin (Costa et al. 1997a).
Further details about the high energy characteristics of this burst
will be published at a later time (Costa et al. 1997b). The burst
localization was further reduced by combining the intersection of the
WFC error circle with a timing arc, obtained (Hurley et al. 1997a,b)
from burst arrival times between the Ulysses and Compton Gamma Ray
Observatory spacecraft. We refer to this error quadrilateral as the
BeppoSAX-IPN (Interplanetary Network) error box. Following this, a
re-calibration of the WFC by in 't Zand et al. (1997) resulted in a
much-reduced error circle for \grb. This new BeppoSAX error circle
remains consistent with the Hurley et al. (1997b) IPN but it is 4.2
arcmin from the center of the old BeppoSAX error circle and has a
radius of only 3 arcmin. In Figure 1 we present a radio image of this
entire region with the different error boxes marked to aid in the
subsequent discussion.

%Follow-up observations 16 hours later with the narrow field
%instruments on board BeppoSAX found two X-ray sources, J1528.8+1944
%and J1528.8+1937 (Butler et al. 1997), interior to the 10 arcmin
%error circle with f$_x\sim{2-3}\times{10}^{-13}$ erg cm$^{-2}$
%s$^{-1}$ between 2-10 keV.  The ROSAT All Sky Survey imaged this field
%in August 1991 and detected X-ray sources at comparable brightnesses
%at these locations (Voges, Boller \& Greiner 1997). J1528.8+1944 is
%the sole X-ray source inside the BeppoSAX-IPN error box.

The radio observations were conducted at the Very Large Array (VLA),
with the first image of the \grb\ field made on January 12.58, 1997
UT, only 28 h after the detection of the gamma-ray burst. This was
followed by a regular monitoring program of the full 10 arcmin radius
WFC error circle at a radio frequency of 1.43 GHz. Table 1 lists the
details of the observation. Dates are given for the start of each
radio observations (Epoch), as are the time elapsed since the
gamma-ray burst ($\Delta$t), the synthesized beam, the rms noise in
each image ($\sigma_{rms}$) and the peak flux density (S$_{1.4}$)
measured for VLA J1528.7+1945 (see \S{3} for details).  The data were
reduced in the usual manner using the AIPS software package.  
%The
%primary flux calibrator used was the radio source 3C\thinspace{286}
%and the phase calibrator was 1513+236. The bootstrap fluxes of
%1513+236, as determined from 3C\thinspace{286}, remained constant
%throughout this period at 1.67$\pm0.01$ Jy and 1.72$\pm0.01$ Jy for
%observing frequencies of 1464.9 MHz and 1385.1 MHz, respectively.

\section{Results}

Fig. 1 is a radio continuum image at 1.4 GHz towards \grb\ taken
almost two years earlier, in March 1995, as part of the NRAO VLA Sky
Survey (NVSS) (Condon et al. 1997). This is the full field of view
covered by our 1997 monitor program of \grb. We have superimposed the
different IPN and WFC error regions discussed above. Although the
synthesized beamsizes are very different, we find all of the NVSS
sources in our monitor data. The majority of these sources are
partially or fully resolved at our higher angular resolution. The NVSS
survey is complete above 2.5 mJy (the second contour), and so with our
better sensitivity we also detect fainter sources (S$>$0.5 mJy) with
no NVSS counterpart.

%All radio sources
%detected by our current VLA observations are indicated by crosses,
%with the largest crosses for bright sources (S$>$10 mJy), the smallest
%crosses for the weakest sources (S$<$1 mJy), and intermediate sizes
%for 1~mJy$\leq{\rm{S}}\leq{10}$~mJy. 

The best current error box for \grb\ is a trapezoidal region defined
by the intersection of the refined IPN annulus (Hurley et al. 1997b)
and the new WFC error circle (in 't Zand et al. 1997). Interior to
this region we detect no radio sources. While our rms noise limits
vary from one day to the next, we can confidently rule out either a
constant or time-variable radio source in the error box above 0.5 mJy
at $\lambda$=20 cm over the time period covered in Table 1. This
corresponds to limits between 3$\sigma$ and 8$\sigma$, depending on
the integration time used on a given day. Galama et al. (1997) report
the detection of a single radio source in the trapezoidal region with
a flux density of 0.6 mJy at a frequency of 840 MHz. We have searched
our images at this position over the same interval that Galama et
al. (1997) claim a detection and we see no such source. In order for
this source to have been detected by the Westerbork Synthesis Radio
Telescope (WSRT) at 840 MHz but remain undetected at the VLA at 1.4
GHz its spectra index would have to be $-2.0$, a value that is rarely
seen except for pulsars.  Alternatively, given the large synthesized
beam size of WSRT, the source may be erroneous, caused by confusion
from a background of unresolved radio sources.

Four radio sources were seen within the larger BeppoSAX-IPN error box
but outside the trapezoidal region. The brightest source, VLA
J1528.7+1949 was detected in the NVSS. The remaining three sources
have average flux densities between 0.4 and 1.4 mJy and are
uncatalogued radio sources. From radio source counts (White et
al. 1997) we expect 0.025 radio sources per arcmin$^2$ with flux
density above 1 mJy, or $2\pm 1$ sources in the BeppoSAX-IPN error
box. Thus the detection of four sources is consistent with all these
sources being unrelated background radio sources. Two of these, VLA
J1528.2+1933 and VLA J1528.6+1940, lie 3.7 arcmin and 4.7 arcmin
from the center of the trapezoidal error box.

An X-ray source SAX J1528.8+1944 (Butler et al. 1997) has been noted
to coincide, within its localization error of 1-arcmin (radius), with
one of the radio sources, VLA J1528.7+1945 (Frail et al. 1997). This
source was first detected by the X-ray satellite ROSAT in August 1991
(Voges, Boller \& Greiner 1997) and was recently re-detected with the
main focal plane cameras aboard BeppoSAX. With the exception of the
source VLA J1528.7+1945 the remaining three radio sources in the
BeppoSAX-IPN error box are steady sources, as are most background
extragalactic radio sources on timescales of days to weeks. A $\chi^2$
test for radio variability (Machalski \& Magdziarz 1993) was carried
out on all eight point sources in the original 10 arcmin BeppoSAX
error circle.  J1528.7+1945 is the only variable source in the field,
with a confidence level which exceeds 99.95\%. The light curve for VLA
J1528.7+1945 is shown in Fig. 2 with several comparison sources. In
March 1995 it was undetected above the NVSS limit of 2.5 mJy. Within
three to four days after the burst it appears to peak at 2 mJy and
decayed to half of this value some 15 days later, where it has
remained level since.  We will discuss this source in more detail in
\S{4}.

\section{Discussion}

\subsection{Constraints on Cosmological Models}

We have imaged a 625 arcmin$^{2}$ region around the position of \grb\
with the VLA on timescales of days to $\sim$months after the initial
burst. We have found no radio sources within the $\sim$14 arcmin$^{2}$
trapezoidal error box, defined by the intersection of the refined WFC
error circle and the IPN. A conservative limit on any flaring or
fading radio counterpart to \grb\ at $\lambda=20$ cm on these
timescales is 0.5 mJy. The present VLA observations provide strong
constraints on the predictions of late-time emission from cosmological
models of GRBs.

The fast risetimes and the energetics of GRBs make it inevitable that
GRBs involve a burst of relativistic particles and magnetic fields.
The sudden deposition of this energy creates a ``fireball'' whose
physics has been studied by various authors (M\'esz\'aros 1995).  A
generic prediction of these studies is that the gamma-ray burst is
followed by bursts at smaller photon energies but with burst
timescales much longer than that at gamma-ray energies and
increasingly delayed with respect to the gamma-ray outburst.  In
models in which the GRBs are of Galactic origin the fireball energy is
relatively small and the expectation is that the radio outburst will
be essentially prompt. However, in cosmological models, the radio
outburst is predicted to be delayed by days to weeks with similar
values for the outburst timescales.

The Paczy\'nski \& Rhoads (1993) model is a variant on the popular van
der Laan (1966) model describing the flux evolution of expanding radio
sources taken to its relativistic limit. Some fraction of the bulk
kinetic energy of the fireball from the GRB goes into magnetic fields
and a relativistic electron population with a power-law distribution
of energies. Baryons account for the majority of energy in the
fireball, the energy density in electrons and magnetic field is
assumed to be less than 1\%. Expressions for the peak flux density and
time-to-maximum of the radio transients predicted in the Paczy\'nski
\& Rhoads model at $\lambda=20$ cm are given in Frail et al. (1994)
for GRB\thinspace{940301}. Since the fluences of the two GRBs were
comparable, we expect similar peak flux densities and timescales of
3.6 mJy and 28 days, respectively. A time-variable source of this
magnitude and on this timescale would have been readily detectable by
our experiment. The above limits are based on the nominal distance,
ambient density and efficiencies assumed by Paczy\'nski \& Rhoads
(1993) and can be relaxed somewhat. For example, the ambient density
was assumed to be 10$^{-24}$ g cm$^{-3}$. Lowering this to a more
reasonable 10$^{-27}$ g cm$^{-3}$ shortens the timescale ($\sim$1 day)
and reduces the peak flux (1.5 mJy) but again such a fading source
would also have been detected.

M\'esz\'aros \& Rees (1977) have also calculated the emergent spectrum
at longer wavelengths at late times. The fireball, formed during the
GRB, interacts with the surrounding medium and dissipates its energy
in part through shocks. The shocks accelerate particles, which in turn
radiate via the inverse Compton and/or synchrotron processes. The
emission peaks at a frequency $\nu_m$ and is self-absorbed below
$\nu_{ab}$. In time the two frequencies decrease as the fireball
decelerates, eventually becoming equal (usually) in the radio regime.
The peak flux density and time-to-peak depend not only on the burst
fluence and the source distance but also on whether the GRB energy is
released impulsively, or in a wind, and on the type of shocks, and the
strength of the B-field. Two impulsive models (their models A2 and A3)
are the most effective at accelerating particles which radiate in the
radio regime. We estimate peak flux densities at 1.4 GHz between 1.7
and 8 mJy on timescales of 40 to 50 days after the burst. The
remaining impulsive and wind models do not predict detectable radio
emission (i.e. $<100~\mu$Jy) at this frequency.

\subsection{VLA J1528.7+1945: An Unusual Radio Source}

Most of the faint radio sources in the sky are thought to be distant
galaxies whose radio emission arises either as a consequence of an
active nucleus (with or without extended radio lobes) or from the
combination of thermal and synchrotron radio emission which
accompanies star formation. Thus the constancy of the flux on short
timescales is understandable. Variations are seen on longer
timescales.  At meter wavelengths, corresponding to radio frequencies
less than a GHz, variations are induced by multipath propagation
through the interstellar medium (Mitchell et al. 1994). At frequencies
above a GHz, the variations are due to intrinsic changes in the
nuclear radio source.  As the frequency increases the amplitude of
these intrinsic variations (Gregory \& Taylor 1986, Machalski \&
Magdziarz 1993) increase and their timescales decrease. Viewed in this
context, VLA J1528.7+1945 is an unusual source because it shows strong
variations on short timescales (Figure 2) at a relatively low
frequency of 1.4 GHz.

We now evaluate the uniqueness of the source VLA J1528.7+1945.  The
study of variability in radio sources, especially faint radio sources,
on short timescales of days and weeks is still in its infancy. The
FIRST survey is a large radio survey of the North Polar Cap, now
partially complete (White et al. 1997). The FIRST images have a
limiting sensitivity and angular resolution nearly identical to those
of our monitoring program which makes it ideal to evaluate the
uniqueness (or lack thereof) of our radio source VLA J1528.7+1945.
During the course of the FIRST survey, about 75,000 sources were
monitored on timescales of days to weeks.  A total of fifty sources
which showed variations of 25\% or larger were identified. Of these
only 12 showed variations exceeding a factor of two (Helfand et
al. 1996). We refer to this small subset as the extreme
variables. From this we conclude that the fraction of extreme
variables is $10^{-4}$. Thus the mean expectation of the number of
extreme variables in our sample of four sources is less than
$10^{-3}$. While this extreme variable lies in the BeppoSAX-IPN error
box it is outside the current best WFC localization for \grb, thus it
appears not to be related to the gamma-ray burst. Nevertheless, the
source VLA J1528.7+1945 is a highly unusual radio variable and is
worthy of further study (Kulkarni et al. 1997).

\section{Conclusions}

The present observations of \grb\ are nearly an order of magnitude
deeper than a similar comprehensive monitoring effort on
GRB\thinspace{940301}.  We note that two other GRBs with comparable
gamma-ray fluences, GRB\thinspace{930706} and GRB\thinspace{920501}
were observed with the VLA as early as 9 and 14 days after the burst
(Palmer et al. 1995) and no radio sources were reported in these error
boxes. Continued long-term monitoring of a dozen arcminute-sized error
boxes on timescales of years has also failed to find time-variable
radio sources (Frail \& Kulkarni 1995) (and Frail et al. 1997b, in
preparation). The absence of a flaring/fading radio source at the
milliJansky level on these timescales is contrary to the predictions
of several models for the late-time emission expected from GRBs at
longer wavelengths.

While the detection of a delayed radio counterpart to a GRB would be a
breakthrough in the study of these enigmatic objects, a non-detection
in itself does not pose a serious challenge to the cosmological
model. The underlying particle acceleration mechanisms, their
efficiencies, geometry, etc are sufficiently uncertain that the
cosmological model cannot be rejected on these grounds. The VLA
observations provide the most sensitive limits which are currently
possible at GHz-frequencies and as such they constitute hard
constraints that any model has to obey. Since all known gamma-ray
sources are radio emitters at some level, we remain hopeful that
continued monitoring of bright GRBs will result in the eventual
detection of a radio counterpart. Now that BeppoSAX has decreased the
positional accuracy to $\pm$3 arcmin it has become possible to observe
at radio frequencies above 5 GHz, where {\it all} cosmological models
of GRBs predict that the radio emission peaks sooner and at higher
flux density levels.

\acknowledgments

The VLA is a facility of the National Science Foundation operated
under cooperative agreement by Associated Universities, Inc.  DAF
thanks Jim Condon for his advice on the interpretation of NVSS data.
SRK thanks David Helfand for discussion on time variability of the
FIRST sources. SRK's research is supported by the National Science
Foundation and NASA.

\clearpage
\begin{figure}
\caption{A radio continuum image at 1.4 GHz towards \grb\ taken in 
March 1995, as part of the NRAO VLA Sky Survey (NVSS). The lowest
contour is plotted at three times the rms noise of 0.5 mJy
beam$^{-1}$, for a synthesized beam size of 45$^{\prime\prime}$. The
brightest radio source in the image is 135 mJy. The large 10 arcmin
radius circle is the earlier BeppoSAX localization. It is intersected
by an annulus (two diagonal lines) determined by Hurley et al. (1997b)
from the Ulysses and the Compton Gamma Ray Observatory spacecraft. The
two smaller circles (60$^{\prime\prime}$ radius) mark the positions of
the X-ray sources detected by BeppoSAX and ROSAT. The 3 arcmin radius
circle is the newest BeppoSAX WFC localization (in 't Zand et
al. 1997). The best current error box for \grb\ is a trapezoidal
region defined by the intersection of the refined IPN annulus and the
new WFC error circle. All radio sources detected by our 1997 monitor
program are indicated by crosses, with the largest crosses for bright
sources (S$>$10 mJy), the smallest crosses for the weakest sources
(S$<$1 mJy), and intermediate sizes for
1~mJy$\leq{\rm{S}}\leq{10}$~mJy. All four radio sources in the
BeppoSAX-IPN error box, one of which was detected previously by the
NVSS, are labeled.}

\caption{Radio light curves for J1528.7+1945 and several comparison 
sources in the field. The plotted error bars are the 1$\sigma$
uncertainties determined from the Gaussian post-fit residuals
(typically 0.1 to 0.3 mJy). J1528.7+1945 is the only variable source
in the entire field, with a confidence level which exceeds 99.95\%.}

\end{figure}

\end{document}